\documentstyle[sprocl]{article}

\def\drop#1{}

\def\cut#1{}
\newcommand{\equ}[1]{~Eq.~(\ref{#1})}

\newcommand{\sla}{\raise.15ex\hbox{$/$}\kern -.8em}
\newcommand{\one}{{\bf 1}}
\newcommand{\Torder}{{\bf T}}
\newcommand{\real}{{\bf R}}
\renewcommand{\natural}{{\bf N}}
\newcommand{\half}{\frac 1 2}

\newcommand{\eps}{\varepsilon}\newcommand{\s}{\sigma}
\newcommand{\bra}{\langle}\newcommand{\ket}{\rangle}
\newcommand{\be}[1]{\begin{equation}\label{#1}}
\newcommand{\ee}{\end{equation}}
\newcommand{\ba}[1]{\begin{eqnarray}\label{#1}}
\newcommand{\ea}{\end{eqnarray}}

\newcommand{\sumint}{\sum\hspace{-1.3em}\int}

\title{Quantum Finance}
\author{Martin Schaden$^\dagger$}
\address{New York University, 4 Washington Place, New York, NY 10003}
\date{\today}
\begin{document}
\maketitle
\begin{abstract}
\noindent Quantum theory is used to model secondary financial
markets. Contrary to stochastic descriptions, the formalism
emphasizes the importance of trading in determining the value of a
security. All possible realizations of investors holding
securities and cash is taken as the basis of the Hilbert space of
market states. The temporal evolution of an isolated market is
unitary in this space. Linear operators representing basic
financial transactions such as cash transfer and the buying or
selling of securities are constructed and simple model
Hamiltonians that generate the temporal evolution due to cash
flows and the trading of securities are proposed. The Hamiltonian
describing financial transactions becomes local when the
profit/loss from trading is small compared to the turnover. This
approximation may describe a highly liquid and efficient stock
market. The lognormal probability distribution for the price of a
stock with a variance that is proportional to the elapsed time is
reproduced for an equilibrium market. The asymptotic volatility of
a stock in this case is related to the long-term probability that
it is traded.

\end{abstract}
\footnotetext{$^\dagger$ Email address: m.schaden@att.net}
\section{Introduction}
Modern quantitative finance is driven by stochastic models of
market behavior. In conjunction with arbitrage arguments these
models provide insights into price relations\cite{BS73,Hull97},
notably between the prices of derivatives and their underlying
assets. However, the stochastic description all but ignores the
cause for market fluctuations, which to a large extent arise
because the price of a security is newly negotiated every time it
is traded\cite{Fama65}. Even rather sophisticated stochastic
models are designed to simulate the observed equilibrium
fluctuations\cite{Baaquie97,Stanley95}. Coherent effects from
trading usually are not incorporated. The difficulties such models
face in describing phenomena like those occurring in a sell-off,
can be traced to the implicit assumption that the trade of a
security does not alter the price distribution for the next trade.
That coherent effects do exist becomes apparent when one tries to
estimate the probable wealth of a majority shareholder. At any
given time, the average worth of a security held by the majority
investor generally may not even be close to the average price
found by trading small amounts of the security: an attempt to
trade a large portion of the majority holdings will almost
invariably change the average price of the security significantly.

We will consider the trade of a security as the basic process that
measures its momentary price. The fact that such a measurement can
only be performed by changing the owner of the security fits the
Copenhagen characterization\cite{Bohr} of a quantum system rather
well. According to this interpretation, the essence of a quantum
system is that any measurement alters its state by a finite
amount. At the quantum level, a measurement may change the outcome
of subsequent measurements in a manner that is not described by
ordinary probability theory. By explicitly describing the
transition process, quantum theory also provides an understanding
of coherent macroscopic effects.

An outstanding manifestation of coherent quantum effects is the
physical phenomenon that occurs in lasers. Some materials under
certain conditions emit light coherently. This can be understood
and becomes predictable only in the framework of quantum theory.
The best statistical description of the behavior of isolated atoms
does not indicate under what circumstances a collection of like
atoms will lase or not. Stimulated emission is a quantum
phenomenon. A similar ability of understanding and predicting the
conditions for collective financial effects is desirable.

Perhaps the best known quantum phenomena observed in physical
systems are associated with interference.  However, these delicate
effects are best observed in relatively simple and controlled
situations.  Compelling evidence for interference phenomena in
finance is still lacking, but this may be due to the complexity of
most financial environments and the lack of controlled
experimentation that could uncover such effects. Some evidence for
financial interference is provided by the technical analysis of
stock prices and the observation of recurrent patterns in large
samples of historical data. Although the recognition of such
patterns is of evanescent financial value due to adjustments in
the market that occur once the pattern is exploited, the
statistically significant and recurrent existence of any pattern
{\it before} its exploitation, could already be considered
evidence for wave-like financial phenomena. There also are some
theoretical arguments for the existence of periodic- and
associated wave- phenomena in finance. All economic developments
after all occur within the confines of the calendar. Agricultural
commodities depend on seasonal effects, and the financial
reporting and taxation of most corporations is quarterly as well.
It is difficult to imagine how such periodic pulses of information
should not give rise to periodic variations of cash flows in
secondary markets. The time required for the dissemination of new
information and for investors to act on it makes wave-like
propagation of at least some financial variables likely.

Although of great interest in the verification of a quantum
description, empirical evidence of interference, diffraction and
other wave phenomena in finance will not be presented here.
Without a fundamental model that predicts the recurrence of
distinct patterns, evidence obtained from a sophisticated
statistical analysis of market data at best can be circumstantial
and probably is inconclusive. At this stage, it is impossible to
decide whether a quantum description of finance is fundamentally
more appropriate than a stochastic one, but quantum theory may
well provide a simpler and more effective means of capturing some
of the observed correlations.

The specific question we will examine here is whether the temporal
evolution of the probability distribution of simple stochastic
models could be the result of a quantum model (for quantum
mechanics as a diffusion process, see Ref.[7]).\footnote{We here
do not emulate diffusion and develop a quantum model of finance
that reproduces stochastic results in certain limits only. The two
descriptions generally are not equivalent\cite{Bell}.} Under
certain equilibrium conditions, the quantum model of financial
markets developed below closely resembles a random walk. However,
the quantum interpretation is based on a detailed description of
the trading process. Microscopically, it is a time reversible
model. The observed dissipation is due to the large number of
rather similar investors. A small and isolated market with only a
few participants would show a very different and far more periodic
behavior.

In quantum theory, the constraints that the probability of any
particular outcome is non-negative at all times and that some
outcome is certain are in a sense trivialized(for a more axiomatic
approach, see Ref.[8]).  To illustrate this consider a stochastic
model for the temporal evolution of the probabilities $w_i(t)$ for
a discrete set of possible events $i=0,1,2\dots$ to occur at time
$t$.  The stochastic process has to ensure that,
\be{constr}
w_i(t)\ge 0,\ \forall\,i  \ \ {\rm and}\ \sum_i w_i(t)=1\ ,
\ee
at all times $t$. The temporal evolution of the probabilities
$w_i(t)$ can, for instance, be described by transition
probabilities $P_{ij}(t',t)$, that give the conditional
probability that the system will be in state $i$ at time $t'>t$,
if it is in state $j$ at time $t$. Due to\equ{constr}, these
matrices do not form a group. In quantum theory the probabilities
are the squares of the moduli of complex amplitudes $A_i(t)$,
\be{amplitudes}
w_i(t)=|A_i(t)|^2,
\ee
and the first set of constraints in\equ{constr} is automatically
satisfied for any choice of amplitudes. That some outcome is
certain requires that the amplitudes satisfy,
\be{uni}
\sum_i |A_i(t)|^2=1=\sum_i w_i(t),
\ee
at all times. \equ{uni} states that the $A_i$ lie on a (not
necessarily finite dimensional) hypersphere. The normalization
condition in\equ{uni} does not depend on time if the amplitudes at
different times are related by a (also not necessarily finite
dimensional) unitary matrix $U_{ij}(t',t)$,
\be{unimat}
A_i(t')=\sum_j U_{ij}(t',t)A_j(t),\ {\rm with} \sum_k
U_{ki}(t',t)U^*_{kj}(t',t)=\delta_{ij}.
\ee
Here $^*$ denotes complex conjugation and Kronecker's
$\delta_{ij}$ symbol stands for the $(ij)$-entries of a unit
matrix. \equ{unimat} guarantees a probability interpretation if
the amplitudes are properly normalized at some particular time
$t_0$. The unitary evolution matrices do form a group
and\equ{unimat} describes the evolution of the amplitudes in
quantum theory.

The relation\equ{amplitudes} between the amplitudes $A_i(t)$ and
the probabilities $w_i(t)$ is not one-to-one. Any two sets of
amplitudes that differ only in their phases correspond to the same
set of probabilities. Stationary probabilities for instance are
described by amplitudes whose phases can be functions of time. Any
conceivable temporal evolution of probabilities can be reproduced
by a temporal evolution of some corresponding set of amplitudes.
However, the converse statement that every unitary evolution of
the amplitudes may be reproduced by a stochastic process for the
probabilities is not true\cite{Bell}.

Whether a quantum model encodes market mechanisms efficiently and
is of greater practical use than a stochastic model largely
depends on the evolution operator. The quantum models below are
not meant to be overly realistic, but perhaps reflect some generic
aspects of quantum finance.

\section{A Hilbert Space Representation of the Market}
The market will be represented by a state in a Hilbert space
${\cal H}$\footnote{This is an in general complex linear space
with a scalar product. Dirac's bra-ket notation\cite{QM} is used
throughout. The scalar product of two states is written
$\bra\phi|\psi\ket=\bra\psi|\phi\ket^*$. Note that dual states
$\bra\phi|$ denote linear functionals on ${\cal H}$. Linear
operators $\hat O$ always act on the state to the right and the
hermitian conjugate operator $\hat O^\dagger$ is defined
by:$\bra\phi|\hat O^\dagger|\psi\ket=\bra\psi|\hat O|\phi\ket^*$,
for any $|\phi\ket$ and $|\psi\ket$. An operator $\hat O$ is
hermitian when $\hat O^\dagger=\hat O$.} A basis for this Hilbert
space are the pure states that reflect all {\it in principle}
simultaneously measurable quantities that describe the market.
These states form a basis of ${\cal H}$ because an idealized
measurement could in principle determine that the market is
represented by only one of these states and no other at a given
moment in time.

The basis states correspond to possible events of the probability
theory. Greatly simplified, the market consists of securities of
types $i=1,2\dots,I$, and participants $j=1,\dots,J$ where $I,J$
are integers (often rather large). In addition, participant $j$
has cash credit/debt $x^j$. The latter may have been loaned from
or to other market participants  or have been accumulated by
trading securities. A {\it completely} measurable basis for the
Hilbert space of the market in this case would be the collection
of pure states
\be{basis}
{\cal B}:=\{ |\{x^j,\{n_i^j(s)\ge 0, i=1,\dots,I \}, j=1,\dots, J
\}\ket \}\ .
\ee
\be{number}
 N_i^j(s)=\int_0^s \!ds' n_i^j(s')\ \ ,
\ee
is the integer number of contracts of security $i$ with a price of
less than $s$ dollars that are held by investor $j$. The number of
securities, with a price between $s$ and $s+\eps$ dollars is a
non-negative integer (for any $\eps>0$)\footnote{Since $\eps$ can
be taken arbitrarily small, the number densities $n_i^j(s)$ are of
the form, $n_i^j(s)=\sum_l m_i^j(s_l)\delta(s-s_l)$, where the
$m_i^j(s_l)\in \natural$ are the number of securities with a price
of $s_l$ dollars and $\delta(x)$ is Dirac's
$\delta$-distribution\cite{QM}.}. The $N_i^j(s)$ thus are
non-negative, monotonic and piecewise constant functions with
integer steps. No a priori restrictions on the liquidity $x^j$ of
each investor has been imposed. One possibility of dynamically
implementing such restrictions is given in Section~4.1\ .

In a market represented by a single basis state of\equ{basis} the
worth of every security and of every investor is known. This is
the {\it maximum} amount of information one can possibly have of
the market at any moment in time. A {\it complete} measurement of
the market would entail that all market participants disclose
their cash possessions and trade all securities simultaneously,
thus fixing the price of each single security. The worth and
holdings in each security of every investor at that point in time
are then known with certainty and the market is described by a
single state of the basis ${\cal B}$. Such a {\it complete}
measurement evidently is not practicable and our knowledge of the
market state in reality will never be so precise. Of importance
here is only that a complete measurement could in principle occur,
and that one cannot have (or does not wish to have) an even more
accurate description of the market than that provided by such an
idealized measurement.

At a given moment in time, the market is represented by a state
$|M\ket$, which in general will be a linear superposition of basis
states $|n\ket\in {\cal B}$,
\be{super}
|M\ket=\sumint_n A_n |n\ket,
\ee
with complex amplitudes $A_n$, whose squared modulus $w_n=|A_n|^2$
is interpreted as the probability that the market is described by
the pure state $|n\ket$. We refrain from interpreting the {\it
phases} of the complex amplitudes $A_n$ at this point, although
these phases of course are at the very heart of the quantum
description and could lead to the coherent financial effects one
would like to describe.

For any two states $|m\ket,|n\ket \in {\cal B}$, the assumption
that the basis ${\cal B}$ consists of completely measurable states
implies that,
\be{ortho}
\bra m|n\ket=0,\ {\rm if}\ m\neq n ,
\ee
where $\bra m|$ denotes the state that is dual to $|m\ket$ and
$\bra\cdot|\cdot\ket$ is the scalar product of ${\cal H}$. In
financial terms\equ{ortho} states that if the market is
represented by a pure state $|n\ket\in{\cal B}$, the probability
that it is described by any other state of the basis vanishes.

We define an {\it isolated} market as one in which no new types of
securities are issued nor old ones retracted and whose
participants do not change. The restriction is not as severe as
may appear at first. The time during which market participants and
securities do not change can be extended considerably by allowing
for artificial market participants and securities that do not
trade. The operator describing the issue of securities is given in
Appendix~A.

In an isolated market the Hilbert space itself does not depend on
time. A consistent probability interpretation is possible if
\be{unitary}
1= \sumint_n w_n=\sumint_n\, |A_n|^2\, ,
\ee
holds at all times. \equ{unitary} is the analog of\equ{uni} for a
set of probabilities that is not necessarily discrete. The
temporal evolution of any state that represents an isolated market
thus is given by a unitary transformation.

\section{Basic Operations}
Any {\it trade} is a partial measurement and restricts the market
state at that moment to a subspace of ${\cal H}$. Successive
trades generally will exclude a static, time independent, market
state. The market's state thus evolves in time and at least some
of the amplitudes $A_n$ of\equ{super} will depend on time.
However, there is a {\it unique} pure state that does not evolve.
In a market described by,
\be{vac}
|0\ket:=|\{x^j=0,n_i^j(s)=0,\ \forall\ i,j,s\}\ket\ ,
\ee
there are no cash reserves and no securities. It is impossible to
raise cash and buy or sell securities in this case. An isolated
market described by $|0\ket$ is so described forever. Although the
state $|0\ket$ apparently is the apotheosis of any market, it is
the starting point for constructing all other states of the basis
${\cal B}$.

\subsection{Moving Cash}
One of the elementary financial actions that we would like to
represent as a linear operator on ${\cal H}$ is the transfer of
funds to a participant. The unitary operator\footnote{To
distinguish them from simple numbers, operators on ${\cal H}$
carry a hat ($\hat{\ } $), or tilde ($\tilde{\ }$) throughout this
article.},
\ba{cash} \hat c^{\dagger j}(s) &=&\exp(-i s\hat
p_j),\ {\rm with}\ s\ge 0\ {\rm and}\ \hat
p_j=-i\frac{\partial}{\partial x^j}\cr
&=&\exp\left(-s\frac{\partial}{\partial x^j}\right) ,
\ea
raises the amount of cash held by investor $j$ by $s$
dollars\cite{QM}, that is,
\be{verifyc}
\hat c^{\dagger
j}(s)|\{x^1,x^2,\dots,x^j,\dots,x^J\}\ket=|\{x^1,x^2,
\dots,x^j+s,\dots,x^J\}\ket\ .
\ee
Note that the hermitian conjugate operator $\hat c^j(s)=\hat
c^{\dagger j}(-s)$ lowers the cash holdings of participant $j$ by
$s$ dollars. [Possible restrictions on the liquidity of market
participants are discussed in Section~4.1~.] The $\hat c^\dagger$-
and $\hat c$- operators commute with each other and satisfy the
multiplication relation,
\be{multc}
\hat c^{\dagger j} (s)\,\hat c^{\dagger j}(s')=\hat c^{\dagger
j}(s+s'),\ {\rm with}\ \hat c^{\dagger j}(-s)=\hat c^j(s)\ {\rm
and}\ \hat c^j(0)=1\ .
\ee
\equ{verifyc} implies that,
\be{translation}
[\hat c^{\dagger j}(s), \hat x^k] =\hat c^{\dagger j}(s)\hat
x^k-\hat x^k\hat c^{\dagger j}(s)= -s\delta^{jk}\,\hat c^{\dagger
j}(s)\ ,
\ee
where $\hat x^j$ is the operator whose eigenvalue is the cash
holdings of investor $j$. The commutation
relation\equ{translation} identifies\cite{QM} $c^{\dagger j}(s)$
as a translation operator; it translates the coordinate $x$ to
$x+s$.

We next define creation and annihilation operators for securities.

\subsection{Creating and Destroying Securities}
Since more than one security of type $i$ held by investor $j$ may
have exactly the same price (in the extreme, they may for instance
all be worthless), the buying and selling of securities is
represented by bosonic creation and annihilation operators. Let
$\hat a^j_i(s)$ be the annihilation operator that removes one
security of type $i$ for a price of $s$ dollars from the portfolio
of investor $j$ and let the hermitian conjugate operator $\hat
a^{\dagger j}_i(s)$ denote the corresponding creation operator
that adds such a security to $j$'s portfolio. [If a certain type
of security, such as common stock, is only traded in packages or
contracts, the creation and annihilation operators for that
security refer to the smallest entity that is traded, rather than
the individual securities themselves.] The selling (buying) price
is included in the description, because this is a known
characteristic of the security, that could be (and often is)
recorded at the time of purchase or sale. Note that a security's
price changes with time and therefore is {\it not} a conserved
quantity\footnote{The price of a security is not the analog of a
particle's momentum in particle physics. We will see that $\ln(s)$
is analogous to a particle's position.}. Between trades, the price
of a security in general will not be known with certainty. The
value of a security that is being held can only be estimated as
the price one may expect to achieve when it is traded. It will
become apparent in section~4.2 that a security bought at a certain
price in time evolves into a superposition of states representing
various prices, with amplitudes that correspond to the probability
that the security can be sold at that price.

The price paid for a security and the difference to the price
realized from (instantly) selling it again is essential to any
trade. From the point of view of the investor two securities of
the same type are not equivalent if they are sold at different
prices. Depending on the market, it may be easier or more
difficult to sell two securities of the same type at two different
prices, than to sell both at the average price. Thus, although the
overall return to the investor is the same, the two sales are not
equivalent from a dynamic point of view.

The creation and annihilation operators for securities thus
satisfy the commutator algebra,
\ba{algs}
[\hat a^j_i(s), \hat a^{\dagger k}_l(s') ] &=&
s\delta(s-s')\delta^{jk}\delta_{il} \cr [\hat a^j_i(s), \hat
a^k_l(s')]  &=& [\hat a^{\dagger j}_i(s), \hat a^{\dagger
k}_l(s')]=0\ .
\ea
The RHS of\equ{algs} is scale invariant. The reason for this
perhaps slightly unusual normalization of the creation and
annihilation operators in\equ{algs} will soon become apparent.
Note that the creation and annihilation operators so defined are
dimensionless and that worthless securities commute.

Let $|0\ket\in{\cal B}$ denote the (unique) state of\equ{vac} that
describes a market in which none of the investors has any
securities nor cash. Since no securities can be sold, we have that
\be{char0}
\hat a^j_i(s)|0\ket=0,\ \forall\ i,j\ {\rm and}\ s\ge 0\ .
\ee
Any other state of\equ{basis} with an integer number of securities
can formally\footnote{ Operators such as $(\hat a^{\dagger
j}_i(s))^n$ for $n>1$ are ill-defined and most of the states
in\equ{createbasis} have divergent norm. This problem of the
canonical continuum formulation of a quantum field theory is well
known. In the present case it can be avoided by discretizing the
price on a logarithmic scale, i.e. by choosing some (small) $h$
and considering only the set of prices $s\in\{e^{k h} \$; k\in
Z\}$. In this case, the creation and annihilation operators can be
normalized so that the right-hand sides of the commutation
relations of\equ{algs} are Kronecker's function on the integers.
The basis states corresponding to\equ{createbasis} then have
finite norm. Since most people do not care about the pennies when
millions are at stake, the regularized version of the model in
many ways is closer to reality -- with $h$ representing the
desired accuracy of the returns. This discretization does not
conflict with any of the global symmetries discussed below.
However, the additional integer indices would clutter all
expressions to the point of making them almost unreadable without
qualitatively changing the discussion.  Only formal continuum
expressions will therefore be presented.}be constructed by acting
with creation operators on this zero-state, $|0\ket$ ,
\be{createbasis}
|\{x^j, n_i^j(s)\}\ket\propto \prod_{j=1}^J \hat c^{\dagger
j}(x^j) \prod_{i=1}^I \prod_{\{s;m^j_i(s)\in\natural \}}
\left(\hat a^{\dagger j}_i(s)\right)^{m^j_i(s)} |0\ket\ ,
\ee
where
\be{defm}
m^j_i(s)=\lim_{\eps\rightarrow 0_+}\int_s^{s+\eps} ds' n^j_i(s')
\ee
is the integer number of securities of type $i$ of investor $j$
with a (momentary) price between $s$ and $s+\eps$ dollars.
Using\equ{algs} and\equ{translation} one can show that the states
of\equ{createbasis} are eigenstates of the security number density
operators
\be{numberop}
\hat n^j_i(s):=\frac{1}{s}\hat a^{\dagger j}_i(s)\hat a^j_i(s)\ ,
\ee
and of the cash holding operators $\hat x^j$,
\ba{eigenvalues}
\hat n^j_i(s)|\{x^k, n_l^k(s')\}\ket&=& n^j_i(s)|\{x^k,
n_l^k(s')\}\ket\nonumber\\
\hat x^j|\{x^k, n_l^k(s')\}\ket&=&x^j|\{x^k, n_l^k(s')\}\ket\ .
\ea

Since a security normally has to be paid for, buying a security is
not quite the same as simply acquiring one. It therefore is
convenient to consider the combinations,
\be{buy}
\hat b^{\dagger j}_i(s) := \hat a^{\dagger j}_i(s) \hat c^j(s)\
,\qquad\ \hat b^j_i(s) := \hat a^j_i(s) \hat c^{\dagger j}(s)\ ,
\ee
that take the change in cash of investor $j$ into account when he
buys/sells the security.  Because cash changing operators are
unitary, commute with the creation and annihilation operators for
securities and obey the multiplication rule of\equ{multc}, the
composite $\hat b$-operators defined in\equ{buy} satisfy similar
commutation relations as the $\hat a$ and $\hat a^{\dagger}$'s
among themselves. However, unlike the $\hat a$'s  they do not
commute with the cash holding operators $\hat x$. The commutation
relations\equ{algs} and\equ{translation} together with the
definitions in\equ{buy} lead to the following commutation algebra
for the $\hat b$'s,
\ba{algb}
[\hat b^j_i(s), \hat b^{\dagger k}_l(s') ] &=&
s\delta(s-s')\delta^{jk}\delta_{il} \nonumber\cr
 [\hat b^j_i(s),
\hat b^k_l(s')]  &=& [\hat b^{\dagger j}_i(s), \hat b^{\dagger
k}_l(s')]=[\hat b^{\dagger j}_i(s),\hat p^k]= [\hat b^j_i(s),\hat
p^k]=0 \nonumber\cr [\hat b^{\dagger j}_i(s), \hat x^k] &=&
s\delta^{jk}\hat b^{\dagger j}_i(s)\nonumber\cr [\hat b^j_i(s),
\hat x^k] &=& -s\delta^{jk} \hat b^j_i(s)\ .
\ea
The latter two relations have the interpretation that an
investor's account is credited (debited) by $s$ dollars when a
security is sold (bought) for that amount.

\section{Temporal Market Evolution}
We have so far constructed states that describe a market by the
probabilities that certain holdings in cash and securities are
realized\footnote{As mentioned before, this representation is not
unique, as the phases of the coefficients $A_i$ in\equ{super} are
not determined by the probabilities.}. Of greater interest is the
temporal evolution of such a state. The state, $|M\ket_t$, that
represents the market at time $t$, is related to the corresponding
state at a later time $t^\prime$ by an evolution operator $\hat U$
\be{evol}
|M\ket_{t^\prime}=\hat U(t',t)|M\ket_t\ .
\ee
The temporal evolution of all possible market states thus defines
a {\it linear} operator on the Hilbert space. $\hat U(t',t)$
furthermore is unique if one properly defines ${\cal H}$ so that
the kernel of $\hat U(t',t)$ is empty. \equ{evol} in this case
uniquely associates a state $|M\ket_{t'}$ at time $t'$ with a
state $|M\ket_t$ at time $t$. This is an evolution of probability
distributions, since a state in ${\cal H}$ encodes the probability
that the market corresponds to a particular pure state in ${\cal
B}$. However, since different market states $|M\ket_t$ give the
same probability distributions at time $t$, it is possible to
encode past evolution in the relative phases of the $A_i$. The
evolution in\equ{evol} is capable of encoding a non-Markovian
temporal evolution of the probabilities and even
"non-deterministic" evolutions that cannot be simulated by a
stochastic process\cite{Bell}.

The probability interpretation of the expansion coefficients
in\equ{super} together with the assumption that the market is
isolated, (or equivalently, that ${\cal B}$ is a complete basis at
all times), leads to the time-independent constraint
of\equ{unitary}. In terms of the scalar product on ${\cal H}$,
\equ{unitary} is just the normalization condition,
\be{unitarity}
1={_t}\bra M|M\ket_t\ \ \forall\, t\ .
\ee
If\equ{unitarity} holds for a set of market states $\{|M\ket_t\}$
that span ${\cal H}$, then $\hat U(t',t)$ of\equ{evol} necessarily
is invertible and thus a unitary operator,
\be{U}
\hat U^{-1}(t',t)=\hat U^\dagger(t',t)=\hat U(t,t')\ ;\ {\rm
with}\ \hat U(t,t)=\one\ .
\ee
\equ{evol} implies that for any $\eps>0$,
\be{multU}
\hat U(t'+\eps,t)=\hat U(t'+\eps,t')\hat U(t',t)\ .
\ee
If the unitary evolution operator $\hat U(t',t)$ is differentiable
in the vicinity of $t'=t$ one has an associated hermitian
Hamiltonian, $\hat H(t)$, that generates the temporal evolution of
the quantum system.
\be{Hdef}
\hat H(t)=i {\Big|}_{t'=t}\frac{\partial}{\partial t'} \hat
U(t',t)\ .
\ee
Defining the time-ordered product of bosonic operators in the
standard way\cite{QM} and introducing the time ordering symbol
$\Torder$ one in this case finds,
\ba{H}
\hat U(t',t)&=&\Torder\left[\exp -i\int_t^{t'} dt \hat H(t)\right]\nonumber\\
&=& \one -i\int_t^{t'}\!dt_1\,\hat H(t_1)+ (-i)^2\int_t^{t'}dt_1
\int_t^{t_1} dt_2 \hat H(t_1) \hat H(t_2)+\dots\nonumber\\
&=& \sum_{n=0}^\infty \frac{1}{n !}  \Torder\left[-i\int_t^{t'}
d\xi \hat H(\xi)\right]^n \ .
\ea
The expansion in\equ{H} generally may not converge for arbitrary
times. However, if the derivative $\hat H(t)$ defined by\equ{Hdef}
is a bounded operator, one can show that for any given (market)
state $|M\ket_t$, the corrections due to higher order terms
in\equ{H} are negligible for sufficiently short time intervals,
$(t'-t)\sim 0$ and that the state $|M\ket_t$ satisfies the
Schr\"odinger equation\cite{QM},
\be{Schroedinger}
i\frac{\partial}{\partial t}|M\ket_t=\hat H(t)|M\ket_t\ .
\ee
\equ{Schroedinger} implies a continuous evolution of the state
with time. It is valid only if the change in the state becomes
arbitrarily small for $t'\sim t$, which, as we shall see, is not
always true for financial markets\footnote{Below we obtain the
change of the market state due to an instantaneous cash flow --
inspection shows that it does not satisfy\equ{Schroedinger}
because the state changes by a unitary transformation that is
discontinuous in time.}. By allowing for operators $\hat H(t)$
that are proportional to $\delta(t-t_0)$ the definition of $\hat
H(t)$ by\equ{Hdef} can be extended to include the case where the
time evolution is discontinuous at time $t=t_0$. However, higher
order terms in the expansion\equ{H} of the evolution operator in
this case are not negligible even for arbitrarily short time
intervals and\equ{Schroedinger} does not give the temporal
evolution of the state.

\equ{Schroedinger} says that the change of the market state in a
short interval of time at time $t$ is given by the action of $-i
\hat H(t)dt$ on the state. On the other hand, for the state
representing the market to change, investors must exchange cash
and/or securities. The Hamiltonian that generates the temporal
evolution thus represents financial transactions. In a
sufficiently short time $(t'-t)\sim 0$, it is rather unlikely that
more than one transaction occurs. To model the time evolution of
the market state it thus may be sufficient to consider only
primitive transactions that can be viewed as occurring
instantaneously.

\subsection{Cash Flow}
The cash flows to and from an investor can be due to the buying
and selling of securities, income, consumption, or accrued
interest in a money market account. The latter could in principle
be modelled as arising due to investments. Since income,
consumption and earned interest to some extent may be known in
advance with very little uncertainty, it is of some practical and
conceptual interest to answer the question whether {\it
deterministic} changes in cash can be reproduced by a quantum
model and to obtain the Hamiltonian that represents fixed cash
flows.

To simplify, we rule out the possibility of default and first
consider a money market account as a magical box in which the cash
of market participant $j$ grows at a pre-known rate $r^j(t)$. The
solution to the problem of finding the associated quantum
Hamiltonian hinges on a peculiarity of the classical equation for
the growth of $j$'s cash. Hamilton's equations that describe the
classical time evolution of the cash are
\be{hamilton}
\frac{d x(t)}{dt}=\left. \frac{\partial H(x,p;t)}{\partial
p}\right|_{{x=x(t)\atop p=p(t)}}\ ,\ \ \frac{d p(t)}{dt}=-\left.
\frac{\partial H(x,p;t)}{\partial x}\right|_{{x=x(t)\atop
p=p(t)}}\ .
\ee
Here $H(x,p;t)$ (without hat) is Hamilton's function and not an
operator. With $x(t)=x^j(t)$ and $H^j(x^j,p_j;t)=r^j(t)x^j p_j$,
the first equation in\equ{hamilton} gives the deterministic growth
of $j$'s cash with a known instantaneous rate of return $r^j(t)$,
\be{classgrowth}
\frac{d x^j(t)}{dt}=\left. \frac{\partial r^j(t)x^j p_j}{\partial
p_j}\right|_{{x^j=x^j(t)\atop p_j=p_j(t)}}= r^j(t) x^j(t)\ .
\ee
Up to some function $V(\{x^k\})$ that does not depend on the
momenta, the classical Hamiltonian describing fixed income is
unique. A potential $V(\{x^k\})$ could be used to penalize
borrowing or promote minimal holdings, in short, impose
restrictions on the liquidity of the investors.  For simplicity
and clarity of exposition, we will not consider such refinements
of the model in this article and set the potential $V(\{x^k\})=0$
in the following.

Demanding that the quantum analog of the classical Hamiltonian is
a hermitian operator, one is thus led to consider,
\ba{Hc}
\hat H_c(t)&=&\sum_{j=1}^J \hat H^j_c(t)\nonumber\\
\hat H^j_c(t) &=& \frac{r^j(t)}{2}(\hat x^j \hat p_j+\hat p_j\hat
x^j)= r^j(t)(\hat x^j \hat p_j-{\frac i 2}\one)=r^j(t)(\hat p_j
\hat x^j+{\frac i 2}\one)\ ,
\ea
as the quantum Hamiltonian that describes the temporal evolution
of cash accounts with known interest rates. Using standard
methods\cite{QM} one verifies that
\ba{Greenc}
G_c({\bf x}',t';{\bf x},t)&:=&\bra {\bf x}'|\hat U_c(t',t)| {\bf
x}\ket=\bra {\bf x}'|\Torder
\exp[-i\int_t^{t'}\!d\xi \hat H_c(\xi)]|{\bf x}\ket\nonumber\\
&=&\prod_{j=1}^J \delta\left(x^{\prime
j}\exp[-\int_t^{t'}\!\frac{d\xi}{2}
r^j(\xi)]-x^j\exp[\int_t^{t'}\!\frac{d\xi}{2} r^j(\xi)]\right )\ .
\ea
Thus, if the market at time $t$ is described by the state
\be{Mt}
|M\ket_t=\left[\prod_{j=1}^J \int_{-\infty}^\infty dx^j
\varphi^j(x^j,t)\right]|x^1,\dots,x^J\ket\ ,
\ee
then\equ{Greenc} gives the time dependence of the amplitudes
$\varphi^j(x,t)$ as,
\be{phit}
\varphi^j(x,t')=\sqrt{Z^j(t',t)}\varphi^j(x Z^j(t',t),t) \ ,\ {\rm
with}\ Z^j(t',t)=\exp[-\int_t^{t'}\!d\xi r^j(\xi)] \ .
\ee
The probability distribution $P^j(x,t')$ that investor $j$ at time
$t'$ has $x$-dollars of cash if the probability that he had
$x$-dollars at time $t$ was $P^j(x,t)$, thus is
\be{Pt}
P^j(x,t')=|\varphi^j(x,t')|^2= Z^j(t',t)|\varphi^j(x
Z^j(t',t),t)|^2=Z^j(t',t)P^j(x Z^j(t',t),t)\ ,
\ee
as one may expect. There is no difference between the
deterministic classical calculation and the quantum one for this
case. This can be traced to the fact that the classical
Hamiltonian is {\it linear} in the momenta. The equivalence
between classical and quantum finance in fact extends to any
previously specified cash flows that can depend on the holdings at
the time they occur. One may generalize $\hat H_c(t)$ to,
\be{genHc}
\hat H_{\rm cash~flow}(t)=\half\sum_j\left[\phi^j(\{\hat
x^k\},t)\hat p_j+\hat p_j \phi^j(\{\hat x^k\},t)\right],
\ee
and still maintain the equivalence between the classical- and
quantum- descriptions. The classical equation of motion for $x^j$
lets us interpret the function $\phi^j(\{\hat x^k\},t)$
of\equ{genHc} as the external cash flow rate\footnote{A classical
Hamiltonian of the form\equ{genHc} corresponds to vanishing
classical action. It generates a canonical transformation of the
phase space.} at time $t$ to investor $j$. A single, instantaneous
cash flow of $s$ dollars into the account of investor $j$ at time
$t=t_0$ for instance is modelled by $\phi^j(\{\hat
x^k\},t)=s\delta(t-t_0)$. To specify $\hat H_{\rm cash~flow}(t)$
the cash flows $\phi^j(x,t)$ would have to be known with certainty
in advance. Quantum finance will only come into its own when this
is no longer the case and uncertain cash flows arise from trading
securities.

\subsection{Trading Securities}
Let us, for the moment, ignore initial offerings and repurchases
and assume that the number of securities of type $i$ on the
secondary market does not depend on time. [The generalization to
when this is not the case is discussed in Appendix~A.] Thus, if
someone buys a security, someone has to sell the same. The
primitive transaction therefore is that trader $l$ buys a security
of type $i$ for $s$ dollars from investor $k$ and (immediately)
sells it for $s'$ dollars to investor $j$, crediting/debiting the
cash difference $s'-s$ to his own account. Note that the trader
could, but need not, be the seller or buyer of the security. The
operator $\hat H_{\rm Trade}(t)$ that encodes such primitive
security trades is of the generic form
\be{trans1}
\hat H_{\rm Trade}(t)=\sum_{i,j,k,l}\int_0^\infty
\frac{ds}{s}\int_0^\infty\frac{ds'}{s'} f^i_{jk;l}(s',s;t)\hat
c^{\dagger l}(s'-s)\hat b^{\dagger j}_i(s') \hat b^k_i(s)\ .
\ee
$\hat H_{\rm Trade}(t)$ is hermitian if the amplitudes
$f^i_{jk;l}(s',s;t)$ satisfy,
\be{hermH1}
f^i_{jk;l}(s',s;t)=f^{i*}_{kj;l}(s,s';t)\ .
\ee
The amplitude $f^i_{jk;l}(s,s';t)$ is related to the mean rate at
which trader $l$ buys security $i$ from investor $k$ for
$s$-dollars and sells it to investor $j$ for $s'$-dollars. To
further constrain the amplitudes, one either requires empirical
data on transaction rates or additional assumptions or both. The
following four assumptions of decreasing generality are useful in
constraining the amplitudes.
\begin{itemize}
\item[(A1)] Invariance of the dynamics under global re-scaling of
all prices. It assumes that the market dynamics would be the same
if {\it all} prices were stated in euros instead of in dollars.
The coefficients in\equ{trans1} in this case are functions of
$s'/s$ only. Introducing,
\be{defy}
\nu:=\ln(s'/s)\ ,
\ee
\equ{hermH1} simplifies to,
\be{hermH2}
f^i_{jk;l}(\nu;t)=f^{i*}_{kj;l}(-\nu;t)\ .
\ee
\item[(A2)] If the market is {\it efficient}, every participant
may be expected to have the same return. There should be no
opportunities that can be better exploited by one investor than by
another. Although clearly an idealization, it is widely believed
that low-cost and high-speed electronic trading together with
readily accessible public information tends to improve market
efficiency. The expected wealth $\overline{W}^j(t)$ of investor
$j$ at time $t$ in the present model is the expectation value of
the operator,
\be{wealthop}
\hat W^j=\hat x^j+\sum_{i=1}^I\int_0^\infty\!\! ds\, \hat
b^{\dagger j}_i(s)\hat b^j_i(s)\ ,
\ee
in the (normalized) state $|M\ket_t$ that describes the market at
time $t$,
\be{wealth}
\overline{W}^j(t)= {_t\bra M|\hat W^j|M\ket_t}\ .
\ee
The two contributions to the expected wealth of an investor are
his expected cash- and security- wealth respectively.  Since the
operator $\hat W^j$ defined in\equ{wealthop} does not explicitly
depend on time, the expected wealth of an investor changes due to
the temporal evolution of the market only. Assuming that this
evolution is sufficiently smooth and using\equ{Schroedinger} one
obtains,
\be{changew}
\frac{d}{dt}\overline{W}^j(t)= {_t\bra M|i[\hat H(t),\hat
W^j]|M\ket_t}\ ,
\ee
which for an efficient market is proportional to the expected
wealth of investor $j$ at that time. The market thus is efficient
if,
\be{efficientmarket}
0= {_t\bra M|i[\hat H(t),\hat W^j]-\bar r(t)\hat W^j|M\ket_t},\
\forall\ j,
\ee
where $\bar r(t)$ is the instantaneous expected return common to
all investors. Note that the expected return $\bar r(t)$ can be
changed by de-, respectively in-flation proportional to the
investors wealth. Since,
\be{inflcomm}
i\left[\half\sum_k (p_k \hat W^k+\hat W^k p_k), \hat
W^j\right]=\hat W^j\ ,
\ee
\equ{efficientmarket} is equivalent to
\be{efficientinflation}
0= {_t\bra M|i[\widetilde H(t),\hat W^j]|M\ket_t},\ \forall\ j,
\ee
where
\be{inflat}
\widetilde H(t)=\hat H(t)+\hat H^{\bar r}_{\rm infl}(t)=\hat H(t)
-\frac{\bar r(t)}{2}\sum_k (p_k \hat W^k +\hat W^k p_k)\ .
\ee
Comparing with\equ{genHc} shows that $\hat H^{\bar r}_{\rm
infl}(t)$ generates a (positive) negative cash flow at time $t$
that is proportional to the investor's wealth, i.e. it simulates
(de-), respectively inflation (or subsidies and taxes).
\equ{efficientinflation} asserts that by introducing an
appropriate inflation rate, the expected instantaneous return of
every investor in an efficient market can be set to zero.

Although there may be many investors,\equ{efficientinflation} can
be satisfied in a variety of ways. The market is efficient
independent of its state only if $\widetilde H(t)$ commutes with
all the $\hat W^j$'s. In this case the wealth of every investor,
when detrended by a common rate, is a strictly conserved quantity.
Not so surprisingly, investors do not trade in this case. For
"strong" market efficiency to hold, $\hat H(t)$, would have to
equal $\hat H^{-\bar r}_{\rm infl}$ up to terms that commute with
all $\hat W^j$'s. It is interesting to note that the security part
of the operator $\hat H^{-\bar r}_{\rm infl}$ is of the
form\equ{trans1} with (diagonal) coefficients,
\be{barf}
i(s'-s)f^i_{jk;l}(s',s;t)=\bar r(t) s
s'\delta_{jl}\delta_{kl}\delta(s'-s)\ ,
\ee
that evidently do not allow for trades between different
investors. The assumption of "strong" market efficiency thus
severely restricts the form of $\hat H(t)$, but would not give a
realistic description of the market.

However,\equ{efficientmarket} is a much weaker condition that also
involves the market state. From\equ{efficientinflation} the market
for instance is momentarily efficient, if $|M\ket_t$ is any
eigenstate of $\widetilde H(t)$. More restrictive is the
requirement that if\equ{efficientmarket} is valid at some time
$t$, $\hat H(t)$ of a truly efficient market must be such
that\equ{efficientmarket} continues to hold at later times. An
evolution operator $\hat H(t)$ that is totally symmetric with
respect to the interchange of investors is sufficient to guarantee
this. Such perfect market democracy requires that every investor
statistically behaves as any other under the same financial
circumstances.

The stability of the market state under small perturbations away
from an efficient one clearly is of some interest. One would like
to know the conditions on $\hat H(t)$ that ensure that a
marginally efficient market evolves {\it toward} a more efficient
one. The market evolves toward an efficient one if at any time
$t$,
\be{evol0}
\frac{d}{dt}\,\ln {_t\bra M|i[\widetilde H(t),\hat
W^j]|M\ket_t}\leq 0\ ,
\ee
for all $j$. The stability of an efficient market will not be
further analyzed here. However, the following considerations lead
to a perturbative expansion about an evolution operator that is
efficient in the strong sense.

\item[(A3)] Neither worthless nor
infinitely expensive securities are ever traded. The amplitudes
$f^i_{jk;l}(\nu;t)$ therefore vanish for $\nu\rightarrow
\pm\infty$. In many markets the possible profit or loss on a vast
majority of trades furthermore is incremental. The  amplitudes
$f^i_{jk;l}(\nu;t)$ in this case are sharply peaked about $\nu=0$
and it may suffice to approximate $f^i_{jk;l}(\nu;t)$ by its first
few Fourier moments. Neglecting all moments except the first two
one has,
\ba{local}
f^i_{jk;l}(s',s;t)&\sim& \delta(\nu)[A^i_{jk;l}(t) +\frac{1}{i\nu}
B^i_{jk;l}(t)]\nonumber\\
&=& s\delta(s'-s)[A^i_{jk;l}(t) -\frac{is}{s'-s} B^i_{jk;l}(t)]\ .
\ea
Note that\equ{hermH2} implies that $A^i_{..;l}$ as well as
$B^i_{..;l}$ are hermitian matrices. In the approximation
of\equ{local} the Hamiltonian describing security transactions
becomes {\it local}. Using the definition\equ{cash} of $c^{\dagger
l}(s'-s)$ and\equ{local} one obtains,
\ba{transl}
\hat H_{\rm Trade}^{\rm local}(t)&\sim
&\sum_{i,j,k,l}\int_0^\infty
\frac{ds}{s}\int_0^\infty\frac{ds'}{s'}
s\delta(s'-s)[A^i_{jk;l}(t) -\frac{is}{s'-s}
B^i_{jk;l}(t)]\times\nonumber\\
&&\hspace{2cm} \times[1-i(s'-s)\hat p^l+\dots]\hat b^{\dagger
j}_i(s') \hat b^k_i(s)\nonumber\\
&=&\sum_{i,j,k,l}\int_0^\infty \frac{ds}{s}\hat b^{\dagger
j}_i(s)[A^i_{jk;l}(t)+ s B^i_{jk;l}(t)(i\frac{\partial}{\partial
s}-\hat p^l)]\hat b^k_i(s)\ .
\ea
Introducing the matrices,
\be{matrices}
A^i_{jk}(t):=\sum_l A^i_{jk;l}(t)\ {\rm and}\ B^i_{jk}(t):=\sum_l
B^i_{jk;l}(t)\ ,
\ee
the local Hamiltonian of\equ{transl} has the form,
\ba{Hsimp}
\hat H_{\rm Trade}^{\rm local}(t)&=&\sum_i \hat H^i(t)= \sum_i
[\hat H^i_0(t)+ \hat H^i_{\rm int}(t)]\ {\rm
with}\nonumber\\
\hat H^i_0(t)&=& \sum_{j,k}\int_0^\infty \frac{ds}{s}\hat
b^{\dagger j}_i(s)
[A^i_{jk}(t)+iB^i_{jk}(t)s\frac{\partial}{\partial s}]\hat
b^k_i(s)\ {\rm and}\nonumber\\
\hat H^i_{\rm int}(t)&=&-\sum_{j,k}\int_0^\infty ds\,\hat
b^{\dagger j}_i(s) \left[\sum_l \hat p^l B^i_{jk;l}(t)\right] \hat
b^k_i(s)\ .
\ea
Note that\equ{matrices} relates part of the "free" Hamiltonian,
$\hat H^i_0$, to the interaction $\hat H^i_{\rm int}$.  Comparing
$\hat H_{\rm int}^i$ with the Hamiltonian of\equ{genHc} that
describes the pre-known cash flows of a trader, one is led to
identify the expectation value of $\hat\phi_l(t)$,
\be{dealvalue}
\hat\phi_l(t):=-\sum_{i,j,k}\int_0^\infty ds\hat b^{\dagger
j}_i(s) B^i_{jk;l}(t)\hat b^k_i(s)\ ,
\ee
with the expected cash flow rate into $l$'s account at time $t$.
It evidently is the (average) result from trading securities. Note
that ${_t\bra M|\hat\phi_l(t)|M\ket_t}$ is proportional to the
expected turnover rate of trader $l$ at time $t$ and that the
price of a security does not change while it is traded. Neglecting
price changes and estimating the income of trader $l$ from his
turnover is in keeping with the approximation that the profit or
loss incurred by a trade is small compared to the turnover. Note
that some of $l$'s cash flow may be the result of "self-trading"
(the terms with $j=k$ in\equ{dealvalue}). In this case the cash
flow rate is proportional to the total value of the securities
held by an investor. If the investor is identical with the trader
this term can describe income from investments due to coupons,
dividends etc.

The price of a security changes due to the second term in $H^i_0$.
This term does not involve explicit cash flows. However, due
to\equ{matrices} there is an intimate relation between this term
of $\hat H^i_0$ and $\hat H^i_{\rm int}$: if no cash flows can be
realized in the trade of a security, its price also will not
change. The relation is a consequence of assuming that the trade
of a security conserves cash: the spread goes to the trader.

It perhaps is possible to abstract from the cash holdings of each
individual investor altogether and replace $\sum_l \hat p^l s
B^i_{jk;l}(t)$ by a more general operator $\hat
\Phi^i_{jk}(s,t)=\hat\Phi^{i\dagger}_{jk}(s,t)$ that describes the
interaction due to the exchange of a security of type $i$ at a
price of $s$ dollars between investors $j$ and $k$. Some symmetry
is highly desirable to manage the potential complexity of such an
interaction. The global scale invariance postulated in (A1)
perhaps can be extended to a local (gauge) symmetry\cite{gauge} or
may be part of a conformal invariance.  Possible relations between
the present approach and others based on symmetries will not be
pursued, because the main objective here is the quantization of
finance per se, rather than the determination of the most
appropriate interaction. Let us therefore consider the "free"
zeroth order Hamiltonian $\hat H^i_0(t)$ more closely.

$H^i_0(t)$ is algebraically diagonalized after
Fourier-transformation in $\nu=\ln(s'/s)$. Using,
\be{delta}
s'\delta(s'-s)=\delta(\ln(s'/s))=\int_{-\infty}^\infty
\frac{dq}{2\pi} e^{iq\nu}\ ,
\ee
$\hat H_0^i(t)$ of\equ{Hsimp} can be rewritten in the form,
\ba{H0}
\hat H^i_0(t)&=&\sum_{j,k}\int_0^\infty \frac{ds}{s}\int_0^\infty
\frac{ds'}{s'}\int_{-\infty}^\infty\frac{dq}{2\pi} \hat b^{\dagger
j}_i(s) [A^i_{jk}(t)+iB^i_{jk}(t)s\frac{\partial}{\partial
s}]e^{iq\nu}\hat b^k_i(s')\nonumber\\
&=&\sum_{j,k}\int_0^\infty \frac{ds}{s}\int_0^\infty
\frac{ds'}{s'}\int_{-\infty}^\infty\frac{dq}{2\pi} \hat b^{\dagger
j}_i(s) [A^i_{jk}(t)+q B^i_{jk}(t)]e^{iq\nu}\hat
b^k_i(s')\nonumber\\
&=&\sum_{j,k}\int_{-\infty}^\infty\frac{dq}{2\pi}{\tilde
b}^{\dagger j}_i(q) [A^i_{jk}(t)+q B^i_{jk}(t)]{\tilde b}^k_i(q)\
,
\ea
where ${\tilde b}^j_i(q)$ is the annihilation operator,
\be{defbk}
{\tilde b}^j_i(q):=\int_0^\infty \frac{ds}{s} e^{iq\ln
(s/\lambda)} \hat b^j_i(s)\ ,
\ee
and ${\tilde b}^{\dagger j}_i(q)$ is the hermitian conjugate
creation operator. Note that a change in the arbitrary but fixed
scale $\lambda$ changes the above definition of the creation and
annihilation operators by a $q$-dependent phase that does not
enter $\hat H^i_0$ nor the commutation relations,
\ba{commk}
[{\tilde b}^j_i(q'),{\tilde b}^{\dagger m}_l(q)] &=&
2\pi\delta^{jm}\delta_{il}\delta(q'-q)\nonumber\cr
 [{\tilde b}^j_i(q'),{\tilde b}^m_l(q)]&=&[{\tilde b}^{\dagger
j}_i(q'),{\tilde b}^{\dagger m}_l(q)]=0,
\ea
that follow from the commutation relations in\equ{algb}, and the
definitions\equ{buy} and\equ{defbk}.

Neglecting the interaction part of the Hamiltonian, the possible
"energies" of the security $i$ at time $t$ are the real
eigenvalues of the hermitian $J\times J$ matrix,
\be{EVs}
A^i_{jk}(t)+q B^i_{jk}(t)\ .
\ee
The eigenvalues $E^i_n(q)$ are labeled by the continuous index
$q\in\real$ and the discrete index $n=1,\dots,J$ that identifies
the eigenvalue at $q=0$; $E^i_n(0)$ is an eigenvalue of the matrix
$A^i_{..}$. Since\equ{EVs} is linear in $q\in\real$, the
eigenvalue of a security $i$ increases or decreases asymptotically
with $q$ and is not bounded below or above. This may be an
artifact of the low-$q$ expansion, since the spectrum of $\hat
H_{\rm Trade}$ is bounded if the transition amplitudes
$f^i_{..;l}$ in\equ{trans1} are positive definite. However, even
though securities are quantized as bosons, frequencies that are
not bounded below do not lead to catastrophic effects in the
finite evolution times one is interested in.

The Hamiltonian is further simplified by noting that for vanishing
matrices $B^i_{..;l}$ no gain nor loss is incurred by trading
security $i$. In this case it is reasonable to assume that the
security will not change hands. If the holdings of every investor
in every security stay the same, the time evolution operator
commutes with the number operators for securities of type $i$ held
by each investor. This certainly is the case if $\hat H^i$ can
itself be expressed in terms of the number operators,
\be{ntot}
\hat N_i^j=\int_0^\infty \frac{ds}{s} \hat b^{\dagger j}_i(s)\hat
b^j_i(s)\ ,
\ee
for securities of type $i$ held by investor $j$\footnote{Note that
without being traded, the value of a security cannot be determined
and one therefore cannot be certain that the number of securities
$i$ of a {\it given value} that are held by investor $j$  does not
depend on time.}. If $A^i_{..}$ is a diagonal matrix and
$B^i_{..;l}=0$, $\hat H^i$ becomes a linear combination of the
$\hat N_i^j$,
\be{Hnoprofit}
\hat H^i=\sum_{j=1}^J A^i_{jj}(t) \hat N^j_i\ .
\ee

One is always free to choose the "investors" in security $i$ in
such a way that $A^i_{..}$ is diagonal. This can be viewed as a
way of defining {\it independent} investors in security $i$. In
general such independent investors will be linear combinations of
the original investors, or for that matter, linear combinations of
the independent investors in another security $i'\neq i$. An {\it
independent} investor in security $i'$ need not be independently
investing in security $i\neq i'$. This can be due to common
interests with other investors in security $i$ that he does not
share with regard to security $i'$. An example are the employees
of a company: their investment in securities of their own company
could be much more correlated than in securities of other
enterprises. Bound by agreements and managerial incentives,
employees in extreme cases may be acting as a single independent
investor with regard to their own company, whereas they
individually are independent investors in other securities. The
example also hints that the basis of {\it independent} investors
in a security may depend on time, since some employees could leave
and others join the company over time.

The information needed to determine the {\it independent}
investors in security $i$ is encoded in the $A^i_{jk}$ and it may
seem senseless to diagonalize these matrices. However, it very
often is quite clear which investors are approximately independent
and one can choose the basis accordingly from the outset. Very
helpful in this respect is that the diagonal coefficients
$A^i_{jj}(t)$ in $\hat H$ of\equ{Hnoprofit} can be interpreted as
Lagrange multipliers for the average total number of securities of
type $i$ held by investor $j$ at time $t$. One concludes that if
the holdings of two investors in a security are uncorrelated in
the absence of gain or loss, these investors can be considered
independent. Note that this does not imply that the investor's
holdings are uncorrelated if a profit or loss can be made by
holding the security. Investors in this sense are independent if
they do not share any common interests in a security other than
its profit potential. Independent investors will not invest in a
security due to their political convictions or because their
brother does.

\item[(A4)] In a basis of independent investors, the coefficients
$B^i_{jk;l}(t)$ in\equ{Hsimp} are related to the probability of a
cash flow in the short time interval $dt$ due to the transfer of
security $i$ between (independent) investors $j$ and $k$. If the
security remains in the hands of an independent investor most of
the time, the diagonal elements of the matrices $B^i_{..;l}$
should therefore be much greater than the off-diagonal ones, or
more precisely,
\be{conditionB}
|B^i_{jj;l}(t)|^2\gg \sum_{k\neq j} |B^i_{jk;l}(t)|^2\ ,\ \forall
i,j,l .
\ee
Mathematically, \equ{conditionB} implies that the matrices
$B^i_{..;l}(t)$ are well conditioned and readily inverted
numerically. Financially \equ{conditionB} says that most of the
security's gain or loss in value occurs while an investor is
holding onto it. It precludes the possibility of a large cash flow
while the security is transferred between investors.
\equ{conditionB} is entirely consistent with our previous
assumption that the market allows only incremental trading
profits. The special non-trading case $B^i_{jk;l}(t)=\bar
r(t)\delta_{jl}\delta_{kl}$, considered under assumption (A2),
would guarantee an efficient market independent of the market
state.

If \equ{conditionB} holds, a perturbative expansion in the
off-diagonal elements of $B^i_{..;l}$ should be accurate and
independent investors would remain almost independent in their
decisions even when gains and losses are possible.
\end{itemize}

The assumptions (A1)-(A4) appear to be satisfied on the stock
market during "normal" times. We next derive the probability
distribution for the evolution of stock prices in such an
equilibrium.

\section{The Evolution of Stock Prices in an Equilibrium Market}
In an equilibrium market the conditional probability $P_T(s'|s)$
that a particular stock can be sold for $s'$ dollars if it was
purchased for $s$ dollars a time $T$ ago is known to be close to
lognormal\cite{Hull97} for sufficiently large $T$,
\be{lognormal}
P_T(s'|s)=\frac{1}{s'\s\sqrt{2\pi
T}}\exp\left[-\frac{(\ln(s'/s)-\mu T)^2}{2\s^2 T}\right].
\ee
The parameter $\mu$ in\equ{lognormal} is the expected return of
the stock and $\s$ is known as its volatility. The statistical
interpretation of\equ{lognormal} is that the yield of a stock in
an equilibrium market depends on many {\it additive} and
statistically independent factors, whose number essentially grows
proportional to the elapsed time $T$. Phenomenologically there
appear to be exceptions to this behavior for short time
periods\cite{Baaquie97,Stanley95,Fama88}, during market upheavals
and due to rare events\cite{Hull97}, i.e. the probability that a
stock becomes worthless in a finite time $T$ does not appear to
vanish as rapidly as the distribution of\equ{lognormal} suggests.
However, \equ{lognormal} does seem to reproduce the observed stock
price distribution of "normal" markets on a time-scale of a few
months, especially near the maximum of the distribution, where it
matters most. From this point of view, the hypothesis that all
stock price probability distributions follow\equ{lognormal} after
a sufficiently long time could serve as the {\it definition} of an
equilibrium market.

In the quantum model of finance, one can derive\equ{lognormal}
under the assumption that (A1)-(A4) are satisfied and the behavior
of the market is sufficiently smooth. We show this by calculating
the conditional probability $P_T(s'|s)$ for the price of a generic
stock under such conditions.

Without loss of generality, we consider the stock of company Doe
with index $i=0$. Joe, the investor with index $j=0$, purchases
one contract of Doe stock for $s$ dollars at time $t=t_i$. We are
interested in the probability, $P_T(s'|s)$, that he can sell his
Doe contract for $s'$ dollars at time $t=t_f=t_i+T$. To simplify
the calculation, Joe should have ample cash to finance any trades
during the time of interest. Prior to $t=t_i$ and after $t=t_f$ we
further stipulate that Joe holds no Doe stock.

The state at time $t=t_i$, $|M_i\ket$, that represents Joe's
holding of {\it one} Doe contract worth $s$-dollars is
proportional to,
\be{M0}
|M_i\ket= \hat b^{0\dagger}_0(s)|\widetilde M_i\ket\ {\rm with}\
\hat b^0_0(s)|\widetilde M_i\ket=0\ \forall s.
\ee
At time $t=t_f$ this initial state has evolved to $|M_f\ket$,
\be{MT}
|M_f\ket=\hat
U(t_f,t_i)|M_i\ket=\Torder\exp\left[-i\int_{t_i}^{t_f} \hat
H(t)\right] \hat b^{0\dagger}_0(s)|\widetilde M_i\ket\ ,
\ee
where the time-ordered exponential of operators is defined
by\equ{H}.

The probability that Joe can sell his Doe contract for $s'$
dollars at time $t=t_f$ thus is related to the amplitude,
\be{Green}
G(s',t_f|s, t_i)=\bra\widetilde M_f|\hat
b^0_0(s')|M_f\ket=\bra\widetilde M_f|\hat b^0_0(s'){\bf T}
\exp\left[-i\int_{t_i}^{t_f} \hat H(t)]\right] \hat
b^{0\dagger}_0(s)|\widetilde M_i\ket\ ,
\ee
where $|\widetilde M_i\ket$ and $|\widetilde M_f\ket$ are states
that are annihilated by all the $\hat b^0_0$'s. $\hat b^{\dagger
0}_0(s)|\widetilde M_i\ket$ and $\hat b^{\dagger
0}_0(s')|\widetilde M_f\ket$ are 1-Doe-contract-owned-by-Joe
states.

We assume further that trading is incremental and that the local
approximation of\equ{Hsimp} is sufficiently accurate to describe
the situation. The only change to the previous treatment is that
$\hat H(t)$ includes annihilation and creation operators for Doe
stock and for stock owned by Joe -- the indices $i,j$ for the
types of stock and the investors now range from $0$ to $I$ and
$0$ to $J$ respectively.

Instead of directly computing the amplitude in\equ{Green} let us
first consider the Fourier transformed amplitude,
\ba{GkT}
\widetilde G(q',q;T,t_i):&=&\int_0^\infty \frac{ds'}{s'}
\int_0^\infty \frac{ds}{s} e^{iq'\ln
(s'/\lambda)-iq\ln(s/\lambda)} G(s',t_f|s,
t_i)\nonumber\\
&\sim &\bra\widetilde M_f|{\tilde b}^0_0(q')\exp\left[-iT\hat
H_0(t_i)\right] {\tilde b}^{0\dagger}_0(q)|\widetilde M_i\ket\ .
\ea
In an equilibrium situation, the explicit time dependence of $\hat
H(t)$ should be negligible if the time interval $T=t_f-t_i$ is
short compared to the timescale of fluctuations in the average
market behavior. In addition, the average cash flow rates in and
out of any account are assumed to be small\footnote{At least that
part of the cash flow which is not associated with stock price
changes and therefore cannot be simulated by redefining the
$B^i_{jk}$ coefficients in\equ{H0}}. To leading approximation, the
time dependent operator $\hat H(t)$ of\equ{Green} has therefore
been replaced by the time independent operator $\hat
H_0(t_i)$in\equ{GkT}. It is advantageous to explicitly isolate the
dependence of $\hat H_0(t_i)$ on the creation and annihilation
operators for Doe-contracts-owned-by-Joe. Suppressing the
dependence on the initial time $t_i$ and avoiding a proliferation
of zero-indices by using the simplified notation $A=A^0_{00}(t_i),
B=B^0_{00}(t_i), B_j=B^0_{0j}(t_i), \tilde b(q)=\tilde b^0_0(q),
\tilde b^j(q)=\tilde b^j_0(q) $, $\hat H_0(t_i)$ is decomposed as,
\ba{decomp}
\hat H_0(t_i)&=&\hat H_{{\cal P}}+\hat V +\hat H_{{\cal Q}}\nonumber\\
\hat H_{{\cal P}}&=& \int_{-\infty}^\infty\frac{dq}{2\pi} (A+q B)
\tilde b^\dagger(q)\tilde b(q)\nonumber\\
\hat V &=&\int_{-\infty}^\infty\frac{dq}{2\pi}q  \tilde
b^\dagger(q)\sum_{j=1}^J B_j \tilde b^j(q) \ \ +\ h.c. \,
\ea
with $\hat H_{{\cal Q}}$ denoting the remainder. Note that the
operator $\hat H_{{\cal Q}}$ does not involve $\tilde b(q)$ or
$\tilde b^\dagger(q)$ operators. The generator $\hat H_0(t_i)$
in\equ{H0} is diagonal in Fourier-space and thus
\be{diag}
\widetilde G(q',q;T,t_i) = 2\pi\delta(q'-q) G_0(q;T,t_i)\ .
\ee

To compute $G_0(q;T,t_i)$ exactly, one would have to know {\it
all} the eigenvalues $E_n(q)$ and eigenvectors $X^n(q)$ of the
matrix for Doe-stock of\equ{EVs} as well as the exact initial (or
final) state. We instead assume that the market state $|\widetilde
M_i\ket$ before Joe bought the Doe contract has existed for a
sufficiently long time and is an approximate eigenstate of $\hat
H_Q$ with eigenvalue $E_i$. Even for an equilibrium market, this
assumption may appear to be a simplification that can hardly be
fully justified. Fortunately, the result will not depend
sensitively on the assumed eigenstate and therefore should be
approximately valid even if the market is a superposition of
states with eigenvalues close to $E_i$. In the absence of
Doe-contracts-owned-by-Joe, $|\widetilde M_i\ket$ thus is assumed
to evolve by a simple phase only. To second order in the (small)
matrix elements $B_j$, the excitation $\tilde
b^\dagger(q)|\widetilde M_i\ket$ of this state after Joe has
acquired his Doe contract is an eigenstate with the slightly
shifted eigenvalue\cite{QM} $E_q$,
\be{en2}
E_q=E_i+A+q B -q^2 \lim_{\epsilon\rightarrow 0_+}\sum_Q
\frac{|\bra Q|\sum_{j=1}^J B_j \tilde b^j(q)|\widetilde
M_i\ket|^2}{A+q B+E_i-E_Q-i\epsilon}\ ,
\ee
where the sum extends over eigenstates $|Q\ket$ of $\hat H_Q$,
that is over all states $|Q\ket$ satisfying $\hat
H_Q|Q\ket=E_Q|Q\ket$. With a Hamiltonian of the form\equ{Hsimp}
there are at most as many non-vanishing matrix elements as there
are investors. Without a detailed knowledge of the states and
energies, it is impossible to evaluate the finite sum in\equ{en2}.
However, in an equilibrium market with a large number of investors
in Doe stock, the energies $E_Q$ of many relevant states are
probably themselves rather close to $E_i+A+q B$, the unperturbed
energy of the initial state. Joe, after all, is just one of many
similar investors. In\equ{en2} the initial energy has been given a
small negative imaginary part to ensure that no state grows in
norm for asymptotically large times. This small imaginary part
determines how the case $E_Q=E_i+A+q B$ is to be treated. Since
$\lim_{\epsilon\rightarrow 0_+} {\rm Im}(x-i\epsilon)^{-1}=i
\pi\delta(x)$, the imaginary part of the sum in general does not
vanish. We define real distributions $\alpha(q)$ and $\s^2(q)\ge
0$ by,
\be{sum}
\alpha(q)+i\s^2(q):=\lim_{\eps\rightarrow 0_+}\sum_Q \frac{|\bra
Q|\sum_{j=1}^J B_j \tilde b^j(q)|\widetilde M_i\ket|^2}{A+q
B+E_i-E_Q-i\epsilon}\ .
\ee
For a finite number of investors $J$ and fixed $E_i$, the
imaginary part, $\s^2(q)$, is a sum of $\delta$-distributions with
support at specific $q$-values only. However, this is an artifact
of the assumption that the initial state is an eigenstate with
eigenvalue $E_i$ and of our neglect of higher order contributions
in the local approximation to $\hat H(t)$. Smearing the $\delta$
distributions by taking $\eps$ in\equ{sum} to be small but finite,
results in a function $\s^2(q)\ge 0$ that is analytic near $q=0$.
The required smearing simulates the fact that the energy $E_i$ of
the initial state is not absolutely sharp. Alternatively one could
imagine an even more idealized scenario in which the number of
investors $J$ becomes arbitrary large. The assumption of an
equilibrium market does not require a finite number of investors.
It indeed is difficult to see how the characteristics of such a
market should depend on the number of participants once they are
sufficiently numerous. Letting $J$ tend to infinity and most of
the $B_j$ tend to zero in such a way that\equ{conditionB} remains
valid in many respects is a rather natural point of view. This
limit also can result in a finite density $0\le\s^2(0)\le \infty$.

Because we neglected other corrections to the real part of the
energy of similar magnitude, consistency with the local
approximation demands that we ignore contributions of order $q^2$
to the {\it real} part of the energy $E_q$. However, the {\it
leading} contribution $q^2\s^2(0)$ to the {\it imaginary} part of
the energy should be retained. Note that second order perturbation
theory is sufficient to determine this leading contribution to the
imaginary part -- other approximations differ in order $q^3$ only.
 To order $q$ in the real part and
leading order $q^2$ in the imaginary part, the energy $E_q$ of the
${\tilde b}^\dagger(q)|\widetilde M_i\ket$ state thus becomes,
\be{en22}
E_q=E_i+A+q B -iq^2\s^2\ ,
\ee
where $\s^2$ is formally (note the order of limits) given as,
\be{s2}
\s^2:=\s^2(q=0)=\pi \lim_{\epsilon\rightarrow
0_+}\lim_{J\rightarrow\infty}\sum_Q \delta(A+E_i-E_Q)| \bra
Q|\sum_{j=1}^J B_j \tilde b^j(0)|\widetilde M_i\ket|^2 \ .
\ee

Since the eigenvalues of the original hermitian Hamiltonian
in\equ{Hsimp} are manifestly all real, the origin of a complex
energy in\equ{en22} is worth some discussion. The point is that
the market state $b^\dagger(q)|\widetilde M_i\ket$ in fact is {\it
not} an eigenstate of the Hamiltonian $\hat H_0(t_i)$ even if
$|\widetilde M_i$ is an eigenstate of $\hat H_Q$. The state
$b^\dagger(q)|\widetilde M_i\ket$ for $J\sim\infty$ in general
will be a rather dense superposition of eigenstates of $\hat
H_0(t_i)$. This superposition of many similar but slightly
different phases leads to a decrease in the amplitude of the
$b^\dagger(q)|\widetilde M_i\ket$ state with time that essentially
is never refreshed -- for $q\neq 0$ there are so many
possibilities of decay into other states (by trading the Doe stock
with other investors), that it is less and less likely that Joe
has a Doe stock with slope $q$ after some time. The dense
superposition of similar phases, resulting in a decrease of the
overall amplitude is simulated by the imaginary part of $E_q$
in\equ{en22}. Note that the imaginary part of $E_q$ in\equ{en22}
vanishes at $q=0$ because Joe is an independent investor in Doe
stock whose interaction with other investors is proportional to
$q$.

For normalized states $|\widetilde M_i\ket$, the definitions
in\equ{GkT} and \equ{diag} together with\equ{en22} give,
\be{Gk}
G_0(q;T,t_i)\sim \exp[-i T E_q]=\exp[-q^2 \s^2 T -i q B T -i
(E_i+A) T]\ ,
\ee
where the parameters $\s^2, A, B$, and $E_i$ in general depend on
the initial time $t_i$. \equ{GkT} can be inverted to obtain the
amplitude $G(s',t_f | s,t_i)$ of interest,
\ba{Gs}
G(s',t_f | s,t_i)&=&\int_{-\infty}^\infty\frac{dq}{2\pi}
G_0(q;T=t_f-t_i, t_i)\exp[-i q\ln(s'/s)]\nonumber\\
&\sim &\int_{-\infty}^\infty\frac{dq}{2\pi}\exp[-q^2 \s^2 T -i q
(\ln(s'/s)+B T) -i (E_i+A) T ]\nonumber\\
&=&\frac{e^{-i T(E_i+A)}}{2\s\sqrt{\pi
T}}\exp\left[-\frac{(\ln(s'/s)+B T)^2}{4\s^2 T}\right]\ .
\ea
The joint probability distribution that Joe can buy a Doe contract
for $s$ dollars at time $t_i$ and sell it for $s'$ dollars at time
$t_f$ thus is,
\ba{jprob} P(s',t_f;s,t_i)&\propto &(s's)^{-1} |G(s',t_f |
s,t_i)|^2\nonumber\\
&=& (4\pi \s^2 Ts's)^{-1}\exp\left[-\frac{(\ln(s'/s)+B T)^2}{2\s^2
T}\right]\ .
\ea
The factor $1/(s's)$ between the joint probability and the square
of the magnitude of the amplitude is due to the normalization of
the creation and destruction operators of\equ{buy}.
[\equ{numberop} shows that the selling operator for {\it one}
security with a price between $s$ and $s+ds$ is $\hat
b^j_i(s)/\sqrt{s}$ rather than $\hat b^j_i(s)$.] From\equ{jprob}
one obtains\equ{lognormal} as the (conditional) probability that a
stock bought for $s$ dollars can be sold for $s'$ dollars a time
$T$ later.

Since states with $|q|>1/{\s\sqrt{T}}$ decay rapidly, the
small-$q$ approximation we have used is consistent for
sufficiently long times $T$. Neglected terms of higher order in
$q$ could become important at short times and for stocks (bonds)
with low volatility $\s$. Deviations from the lognormal price
distribution at short times have indeed been
observed\cite{Fama88}. Empirical data\cite{Stanley95} suggests
that $\s^2(|q|\sim\infty)$ scales like $|q|^{-\alpha}$, with an
exponent $\alpha\sim 1$. This behavior would lead to a
symmetrical, L\'evy unstable price distribution at sufficiently
short times $T$. In practice, the lognormal distribution obtained
above, is a good description only for $T>1$month or so.

\section{Discussion}
Perhaps more interesting than the derivation of the lognormal
price distribution for an equilibrium market in the quantum
description are the assumptions that were made and the
interpretation of the parameters of the Hamiltonian this
derivation provides.

\equ{jprob} identifies the diagonal entries of the $B^i_{..}$
matrix with the expected return, $\mu_i$, of the stock in the
equilibrium market,
\be{identB}
B^i_{jj}=-\mu_i,\ \forall\ {\rm investors}\ j.
\ee
The stock's volatility $\s^i$ is given by\equ{s2}. It is
remarkable that to this order of the approximation the "chemical
potentials" $A^i_{jj}$ for the number of securities of type $i$
held by independent investor $j$ enter the volatility only. The
volatility of a stock also depends indirectly on the "state" of
the market, here described by the eigenvalue $E_i$. The expected
return of the stock apparently does not. This is consistent with
the notion that the expected return of a stock is related to the
performance of the company and should not depend on how the
outstanding stock and cash are distributed among investors,
whereas the volatility of a stock does depend on the stock's
distribution among investors as well as on the state of the
market. This separation of effects was not anticipated in the
formulation of the model, which is based on trading probabilities.
\equ{s2} relates the volatility of a stock to the probability that
it is traded. The expression for the volatility in\equ{s2} is
analogous to that of an absorption cross-section and implies that
an increase in the trading rate for a stock should be accompanied
by an increase in its volatility. It is perhaps not unreasonable
that such a relation should exist, since high trading volume
appears to be associated with larger changes in price (on
empirical grounds, a relation between volatility and trading
volume was postulated by Clark\cite{Clark73}). It is, however,
amusing that this highly simplified model of the market gives such
a relation.

Note that we have made extensive use of Fourier-analysis without
financially interpreting the variable $q$, the conjugate quantity
to $\ln (s/\lambda)$. In particle theory $q$ would be the
particle's wave-number -- but what does a wave-number signify
financially? Contrary to the interference demonstrations in
particle physics, no financial experiment presently determines the
wave-number of a stock. The analogy with a quantum particle and
the observation that trading ceases at $q=0$ for any stock,
suggest that $q$ is proportional to the rate of return $q= m
dx/dt+O(q^2)$ for sufficiently small $q$, where the
proportionality constant $m$ is a characteristic of the stock.
[The proportionality between the wave-number and the velocity of a
particle is one of De Broglie's wave-particle duality relations in
the non-relativistic case\cite{QM}.] Heisenberg's famous
uncertainty relation\cite{QM} supports this interpretation. The
uncertainty relation essentially is the mathematical statement
that the variance $\s_x^2$ of a probability distribution
$\rho(x)=|\psi(x)|^2$ and the variance $\s_q^2$ of the
corresponding distribution
$\widetilde\rho(q)=|\widetilde\psi(q)|^2$ for the
Fourier-conjugate variable $q$ with
\be{FT} \widetilde \psi(q):=\int_{-\infty}^{\infty} dx
\psi(x) e^{iqx}\ ,
\ee
satisfy the inequality\cite{QM}
\be{URel}
\s_x^2 \s_q^2\ge {\frac 1 4}.
\ee
\equ{URel} restricts the joint measurement of any two conjugate
variables -- it in particular is not possible to measure both to
arbitrary accuracy. As one distribution becomes more sharply
peaked, the variance of the distribution for the conjugate
variable necessarily increases.

For the lognormal probability distribution of\equ{lognormal}, the
uncertainty in the price of a stock after a period $T$ is $\s_x=\s
\sqrt{T}$. The uncertainty in the (average) rate of return over
this period thus is $\s_{\dot x}\sim \s/\sqrt{T}$.  The standard
deviation of the distribution $\rho(q)$ for the conjugate
variable, which can be read off the Fourier-amplitude (74), is
$\s_q=1/(2\s\sqrt{T})$. The uncertainty relation\equ{URel} in this
case holds as an equality (because the distributions are
Gaussian). The uncertainty of the average rate of return and of
$q$ clearly are proportional with a proportionality constant,
\be{interpretm}
2m=\s^{-2}\ .
\ee
Note that the volatility is an unchanging characteristic of the
stock in the lognormal model (the only one apart from the stock's
average return). The wave-number $q$ and the rate of return $\dot
x$ in general are proportional only near $q=\dot x=0$. The
observed volatility may depend on time and the quantity
in\equ{interpretm} is the volatility for sufficiently long time
intervals (when the distribution is close to lognormal). Similar
to particles in a medium, the observed "mass" $m$ of a stock (or
the inverse of its volatility) furthermore varies with changes in
the environment. It is interesting that equilibration to a
lognormal price distribution after sufficiently long times occurs
only for stocks with $m>0$. This perhaps is related to the
(infrared) instability of 1+1 dimensional field theories with
massless excitations.

Unlike the wave number of particles, that of stocks is not a
financial characteristic that is readily measured or prepared.
However, before the now famous interference experiments, a wave
interpretation of particle behavior would have seemed equally
absurd. Since the amplitudes for conjugate variables are related
by Fourier-transformation, all the information is contained in the
amplitude for either. Monitoring the price of a stock thus
exhausts the available information and no advantage is gained by
also measuring its wave number in some way. Finance in this sense
perhaps is the penultimate quantum model, in which the
construction of the probability distribution for the variable
conjugate to the logarithm of the price is mainly a matter of
mathematical convenience.

Let me finally summarize the assumptions about the stock market,
which in leading approximation lead to the lognormal price
distribution of\equ{lognormal}.
\begin{itemize}
\item[1)] Incremental trading with small profits/losses from
individual trades
\item[2)] A time interval $T$ that is short compared to
characteristic periods of variation of the market as a whole, but
sufficiently long for the low-$q$ approximation to be valid.
\item[3)] The market can be viewed as isolated during the time
interval $T$, without major injections or extractions of cash or
newly issued stock.
\item[4)] In any short period of time it is far more probable that
the stock is retained than that it is traded.
\item[5)] A large number of similar investors in the stock.
\end{itemize}
One should expect deviations from a lognormal price distribution
if one or the other of these condition is not met. Lognormal
distributions in particular probably are not characteristic for a
market with few and/or very different investors or in situations
where relatively large profits/losses in individual trades can
have a significant impact on the price distribution. It may be
worth investigating whether the deviations from the lognormal
distribution observed for short times\cite{Fama88}, which in
stochastic models is simulated by a stochastic
volatility\cite{Baaquie97} or a non-random walk\cite{Stanley95},
in the present formulation require the inclusion of higher
Fourier-moments in the expansion of\equ{local} or simply reflect
the behavior of $\s^2(q)$ for $|q|\sim\infty$ in\equ{sum}.

\noindent {\bf Acknowledgement:} I would like to thank L.~Spruch
and L.~Baulieu for encouragement and helpful suggestions.

\appendix

\section{Non-Isolated Markets, Time Dependent Hamiltonians and the
Generating Functional} The model Hamiltonian\equ{trans1} for
trading preserves the number of securities of every type as well
as the overall available cash. It does not take into account
initial (public) offerings, consumption, or the possibility of
earning cash from other sources. The primary offering of a
security and its initial sale to a group of investors can,
however, be simulated by a coherent initial state. Such a state is
created at time $t=t_0$ by a component of $\hat H(t)$ of the form
\be{IPO}
\hat H_{\rm IPO}(t)= \delta(t-t_0)\sum_j\int_0^\infty \frac{ds}{s}
[\eta^j(s) \hat b^{i\dagger}_j(s)+\eta^{j*}(s)\hat b^i_j(s)],
\ee
if there are no trades of security $i$ for $t\leq t_0$.  The
densities $n^j(s)=|\eta^j(s)|^2/s $ represent the number of newly
issued securities acquired by investor $j$ for a price between $s$
and $s+ds$ dollars, $n^j=\int ds n^j(s)$ being the expected total
number of securities bought by $j$. The amplitudes $\eta^j(s)$
realistically are sharply peaked about the offering price $s_0$.

By generalizing\equ{IPO} to time- and price- dependent sources
$\delta(t-t_0)\eta^j(s)\rightarrow \eta^j(s,t)$ and including
external cash flows described by\equ{genHc}, one can study the
response\cite{QM} of the market to the issue and repurchase of
securities as well as to external income and consumption. In this
formalism, the generating functional ${\cal Z}$ for financial
response functions formally is,
\be{genZ}
{\cal Z}[\eta^j(s,t),\phi^j(t)]:=\bra 0|\Torder
\exp\left[-i\int_{-\infty}^\infty\!dt\, [H(t)+H_{\rm
IPO}(t)+H_{\rm cash~flow}(t)]\right]|0\ket\ ,
\ee
where $\phi^j(t)$ is the rate of cash flow to investor $j$ at time
$t$.

\end{document}